\documentclass[aps,pra,twocolumn,amsmath,showpacs,amssymb,floatfix,nofootinbib]{revtex4}
\usepackage{graphicx,psfrag,bm,times}
\begin{document}
\title{Characterization of two-qubit perfect entanglers}
\author{A. T. Rezakhani }
\email{tayefehr@mehr.sharif.edu}
\affiliation{Department of
Physics, Sharif University of Technology, P. O. Box 11365-9161,
Tehran, Iran}

\begin{abstract}
Here we consider perfect entanglers from another perspective. It
is shown that there are some {\em special} perfect entanglers
which can maximally entangle a {\em full} product basis. We have
explicitly constructed a one-parameter family of such entanglers
together with the proper product basis that they maximally
entangle. This special family of perfect entanglers contains some
well-known operators such as {\textsc{cnot}} and {\textsc{dcnot}},
but {\em not} ${\small{\sqrt{\rm{\textsc{swap}}}}}$. In addition,
it is shown that all perfect entanglers with entangling power
equal to the maximal value, $\frac{2}{9}$, are also special
perfect entanglers. It is proved that the one-parameter family is
the only possible set of special perfect entanglers. Also we
provide an analytic way to implement any arbitrary two-qubit gate,
given a proper special perfect entangler supplemented with
single-qubit gates. Such these gates are shown to provide a
minimum universal gate construction in that just two of them are
necessary and sufficient in implementation of a generic two-qubit
gate.
\end{abstract}
\date{\today}
\maketitle

\section{Introduction}{\label{intro}}

Entanglement has been proved to be a crucial ingredient in many
quantum information processing (QIP) tasks, such as quantum
computation and quantum communication \cite{Nielsenbook}. In this
respect, entanglement is a unique quantum mechanical resource
which its production, quantification, and manipulation are of
paramount importance in QIP. A fundamental relevant question here
is how to characterize entangling capabilities of quantum
operations. In fact, in this regard a lot of investigation have
been done from many different aspects
\cite{Zanardi1,Zanardi2,Ciracdur,Krauscirac,Collins,Durvidal1,Durcirac,Makhlin,Vidalciraccat,Durvidal2,DurciracQIC,
Hammerer,Wangzanardi,Wangsanders,Leifer,Zhang1,Nielsen}.

In this paper we investigate the problem of characterization of
perfect entanglers \cite{Zhang1} from a new perspective. Perfect
entanglers are defined as unitary operators that can generate
maximally entangled states from some suitably chosen separable
states. These operators are, in some sense, half of all nonlocal
two-qubit unitary operators. Thus characterization of their
(geometric and algebraic) structure(s) will be important.
Fortunately, there are some well-defined (and relatively easy to
calculate) tools to measure entangling properties of quantum
operators, and also entanglement properties of two-qubit states
\cite{Zanardi1,Nielsen,Wootters}. By using these tools we show
that one can analyze perfect entanglers and find that some of them
have the unique property of maximally entangling a complete set of
orthonormal product vectors.  We explicitly construct such
operators and show that they have the maximal entangling power,
$\frac{2}{9}$. It is proved that these are the only unitary
operators which have this property. However this investigation has
its own importance, we provide a probable application for them. An
analytic way to use such special perfect entanglers as the
building blocks of the most efficient universal gate simulation is
put forward. This assigns another remarkable role for such special
entanglers in universality issues and the related topics.

The structure of this paper is as follows. In Sec.~\ref{sec2} we
review some necessary definitions and some important facts about
nonlocal two-qubit operators. This section is supplemented with
appendices at the end the paper. In Sec.~\ref{sec3}, we explicitly
construct a one-parameter family of such special perfect
entanglers and some relevant examples are provided. Also we go
further and show that the family we have found is exactly the
family of perfect entanglers with maximal entangling power,
$\frac{2}{9}$, and no other perfect entangler except these ones
have the mentioned property. In Sec.~\ref{sec4} we discuss
applications of these special perfect entanglers in universal gate
simulation. We show that they can be used as elementary building
blocks in minimum universal construction of two-qubit gates. The
paper is concluded in Sec.~\ref{conc}.

\section{Preliminaries}{\label{sec2}}

In this section we want to review briefly some facts on nonlocal
two-qubit unitary operators. More discussion on this matter can be
found in the appendices and references therein. It is known that
\cite{Khaneja, Krauscirac, Zhang1} any two-qubit unitary operator
$U\in$ SU(4) can be written in the following form
\begin{eqnarray}
&\hskip -3mm U=(A_1\otimes
B_1)e^{-i(c_1\sigma_1\otimes\sigma_1+c_2\sigma_2\otimes\sigma_2+c_3\sigma_3\otimes\sigma_3)}(A_2\otimes
B_2),\hskip 2mm \label{canonical}
\end{eqnarray}
where $\sigma_i$'s are the Pauli spin matrices, and $A_i$, $B_i$
$\in$ SU(2) are some single-qubit unitary operators. By using this
decomposition for $U$, called the canonical decomposition, one can
define local equivalence of two unitary operators. Two operators
$U$, $U'$ $\in$ SU(4) are considered locally equivalent, $U\equiv
U'$, if there exist some single-qubit unitary operators
$u_1,u_2,v_1,v_2$ $\in$ SU(2), such that
\begin{eqnarray}
&U'=u_1\otimes v_1 U u_2\otimes v_2. \label{LU}
\end{eqnarray}
It has been shown that using the concept of local equivalence one
can find some freedoms in $c_i$'s  such that one can always
restrict oneself to the region $\frac{\pi}{4}\geq c_1\geq c_2\geq
|c_3|$ in the $(c_1,c_2,c_3)$ space (Appendix \ref{appa}). This is
the so-called Weyl chamber \cite{Zhang1}. Hereafter, whenever we
write $[c_1,c_2,c_3]$ we will assume using these freedoms. So when
we want to deal with local equivalence we can always consider
$U\equiv {\cal U}\equiv [c_1,c_2,c_3]$, up to the freedoms. For
the following uses it is helpful to present the explicit form of
${\cal U}$. In the standard computational basis (where
$\sigma_3|0\rangle=|0\rangle,~\sigma_3|1\rangle=-|1\rangle$), we
have
\begin{eqnarray}
\hskip -3mm&{\cal U}=\left(\begin{array}{cccc} e^{-ic_3}c^{-} & &
&-ie^{-ic_3}s^{-} \\& e^{ic_3}c^{+}&
-ie^{ic_3}s^{+}&\\
& -ie^{ic_3}s^{+}& e^{ic_3}c^{+} &\\
-ie^{-ic_3}s^{-} &&& e^{-ic_3}c^{-}
\end{array}\right),
\label{explicitU}
\end{eqnarray}
where $c^{\pm}=\cos(c_1\pm c_2)$ and $s^{\pm}=\sin(c_1\pm c_2)$.
Consider the usual Bell basis,
$\{|\Phi^{\pm}\rangle=\frac{|00\rangle\pm|11\rangle}{\sqrt{2}},|\Psi^{\pm}\rangle=\frac{|01\rangle\pm|10\rangle}{\sqrt{2}}\}$.
Then, it is easily seen that in the magic basis\footnote{There are
different conventions about the magic basis
\cite{Krauscirac,Hammerer,Wootters}. We follow the definition in
\cite{Krauscirac}.}
\begin{eqnarray}
&|\Phi_1\rangle=|\Phi^{+}\rangle,~~|\Phi_2\rangle=-i|\Phi^{-}\rangle,\\
&|\Phi_3\rangle=|\Psi^{-}\rangle,~~|\Phi_4\rangle=-i|\Psi^{+}\rangle,
\end{eqnarray}
${\cal U}$ is diagonal
\begin{eqnarray}
&{\cal U}=\sum_{k=1}^4e^{-i\lambda_k}|\Phi_k\rangle\langle\Phi_k|,
\label{spectral}
\end{eqnarray}
where
\begin{eqnarray*}
&&\lambda_1=c_1-c_2+c_3,\hskip 7mm \lambda_2=-c_1+c_2+c_3\\
&&\lambda_3=-(c_1+c_2+c_3),\hskip 2mm \lambda_4=c_1+c_2-c_3.
\end{eqnarray*}

Makhlin \cite{Makhlin} and Zhang {\em et al.} \cite{Zhang1} have
shown that the local equivalence classes can be characterized
uniquely by the two local invariants $G_1[U]$ and $G_2[U]$ which
are as follows
\begin{eqnarray}
&&\hskip -4mmG_1=\frac{1}{4}[e^{-2i c_3}\cos 2(c_1-c_2)+e^{2i
c_3}\cos
2(c_1+c_2)]^2,\hskip 4mm\label{invs1}\\
&&\hskip -4mmG_2=\cos (4 c_1)+\cos (4 c_2)+\cos (4 c_3).
\label{invs2}
\end{eqnarray}

Moreover, in \cite{Makhlin,Zhang1} the important notion of a
perfect entangler has been defined. Strictly speaking, a two-qubit
unitary operator is called a perfect entangler if it can produce a
maximally entangled state from a suitable unentangled (separable)
one. It is not hard to see that [{\textsc{cnot}}], [{\textsc{
dcnot}}]\footnote{{\textsc{dcnot}} or double-{\textsc{cnot}} is
defined as \cite{Hammerer}: {\textsc{
dcnot}}=${\rm{\textsc{cnot}}}_{21}{\rm{\textsc{cnot}}}_{12}$,
where the subscripts 12 for {\textsc{cnot}} mean that the first
(second) qubit is the control (target) qubit, and similarly for
21.}, and [${\small{\sqrt{\rm{\textsc{swap}}}}}$] classes are all
perfect entanglers. For example, for [{\textsc{cnot}}] a choice
can be [{\textsc
{cnot}}]$|0\rangle|0\rangle=\frac{|00\rangle-i|11\rangle}{\sqrt{2}}$.

To characterize entangling capabilities of unitary operators
entangling power has been introduced \cite{Zanardi1}. This measure
is defined to be the average entanglement that the unitary
operator can produce when acting on separable states. It can be
shown that entangling power of a unitary operator $U$, $e_p(U)$,
has the following simple form
\begin{eqnarray}
e_p(U)=&e_p({\cal U})=\frac{1}{18}[3-(\cos 4c_1 \cos
4c_2\nonumber\\&+\cos 4c_2 \cos 4c_3+\cos 4c_3 \cos 4c_1)],
\label{EPclosed}
\end{eqnarray}
which only depends on the nonlocal part $[c_1,c_2,c_3]$ (Appendix
\ref{appb}).

 Other important points that are helpful in our next
discussions are separability and maximally entanglement conditions
in the magic basis. It has been shown in \cite{Krauscirac} that a
state $|\Psi\rangle=\sum_k\mu_k|\Phi_k\rangle$ is separable iff
$\sum_k\mu_k^2=0$. As well, $|\Psi\rangle$ is maximally entangled
iff $\mu_k^2=e^{-i\delta}|\mu_k|^2$ ($k=1,\ldots,4$). This means
that a state is maximally entangled iff its coefficients in the
magic basis are real, up to a global phase.

\section{Special perfect entanglers}{\label{sec3}}
In this section, we are going to study some special members of the
space of perfect entanglers. As we see below this turns out to be
useful in characterization of structure of the space of perfect
entanglers. In this regard, we pose the
following question:\\
\indent Does there exist any perfect entangler which can maximally
entangle a {\em full} separable basis?\\
This question is motivated by the following observation which was
firstly reported in \cite{Ours}. To be complete in our discussion
we reproduce it here. It can be seen that for the
[{\textsc{cnot}}] class one can find four orthonormal product
states which are transformed to some maximally entangled states.
For example, we have
\begin{eqnarray}
&[{\rm{\textsc{cnot}}}]|00\rangle=\frac{|00\rangle-i|11\rangle}{\sqrt{2}},
\hskip 2mm
[{\rm{\textsc{cnot}}}]|01\rangle=\frac{|01\rangle-i|10\rangle}{\sqrt{2}},\nonumber\\
&[{\rm{\textsc{cnot}}}]|10\rangle=\frac{|10\rangle-i|01\rangle}{\sqrt{2}},
\hskip 2mm
[{\rm{\textsc{cnot}}}]|11\rangle=\frac{|11\rangle-i|00\rangle}{\sqrt{2}}.
\label{CNOTsep}
\end{eqnarray}
This property may not seem so important at this stage. But if we
want to find such a separable basis for the
$[{\sqrt{\rm{{\textsc{swap}}}}}]$ class the result, as is shown
below, is a failure. To see this, we note that the most general
separable basis (up to general phase factors for each vector) that
can be considered is as follows
\begin{eqnarray}
&&|\Psi_1\rangle=(a|0\rangle+b|1\rangle)\otimes
(c|0\rangle+d|1\rangle),\nonumber\\
&&|\Psi_2\rangle=(-\bar{b}|0\rangle+\bar{a}|1\rangle)\otimes
(c|0\rangle+d|1\rangle),\nonumber\\
&&|\Psi_3\rangle=(e|0\rangle+f|1\rangle)\otimes
(-\bar{d}|0\rangle+\bar{c}|1\rangle),\nonumber\\
&&|\Psi_4\rangle=(-\bar{f}|0\rangle+\bar{e}|1\rangle)\otimes
(-\bar{d}|0\rangle+\bar{c}|1\rangle), \label{separablebasis}
\end{eqnarray}
in which bar means complex conjugation and
$|a|^2+|b|^2=|c|^2+|d|^2=|e|^2+|f|^2=1$. Now if we act on one of
these states, say $|\Psi_1\rangle$, by the [{\textsc{swap}}] class
we obtain
\begin{eqnarray}
&[{\sqrt{\rm{\textsc{swap}}}}]|\Psi_1\rangle=e^{-i\frac{\pi}{8}}ac|00\rangle+\frac{e^{i\frac{\pi}{8}}}{\sqrt{2}}(ad-ibc)|01\rangle
\nonumber\\&-i\frac{e^{i\frac{\pi}{8}}}{\sqrt{2}}(ad+ibc)|10\rangle+e^{-i\frac{\pi}{8}}bd|11\rangle.
\end{eqnarray}
This state is maximally entangled if its concurrence
\cite{Wootters}, $C=|ad-bc|^2$, is equal to $1$ (Appendix
\ref{appb}). Thus the $[\sqrt{\rm{\textsc{swap}}}]$ class turns
these orthonormal states into maximally entangled ones if the
following conditions are fulfilled simultaneously
\begin{eqnarray}
&|ad-bc|^2=1,\hskip 2mm |\bar{a}c+\bar{b}d|^2=1,\label{1stc}\\
&|cf-de|^2=1,\hskip 2mm |\bar{c}e+\bar{d}f|^2=1.\label{2ndc}
\end{eqnarray}
However, by side by side addition of the equations in
(\ref{1stc}), we get $(|a|^2+|b|^2)(|c|^2+|d|^2)=2$ which in the
light of the normalization conditions is an apparent
contradiction! A similar contradiction is obtained from
Eq.~(\ref{2ndc}). Therefore the $[\sqrt{\rm{\textsc{swap}}}]$
class cannot maximally entangle any full product basis. However,
it must be noted that this fact does not forbid finding a pair of
orthonormal states which become maximally entangled by this class.
For example, by choosing the following pair of orthonormal product
vectors
\begin{eqnarray}
&&|\Psi_1\rangle=(a|0\rangle+b|1\rangle)\otimes
(c|0\rangle+d|1\rangle),\nonumber\\
&&|\Psi'_1\rangle=(-\bar{b}|0\rangle+\bar{a}|1\rangle)\otimes
(-\bar{d}|0\rangle+\bar{c}|1\rangle),
\end{eqnarray}
in which $|ad-bc|^2=1$, no contradiction is turned out. It is
interesting to note that
$e_p(\sqrt{\rm{\textsc{swap}}})=\frac{1}{6}$ which is the minimum
entangling power among all perfect entanglers.

In the following we are going to find some other unitary operators
having this property. Hereafter, the perfect entangler that can
maximally entangle a full product basis are called {\em special}
perfect entanglers (SPE). We are trying to explicitly construct
such operators together with their related
product basis.\\
Since working with a general product basis,
Eq.~(\ref{separablebasis}), is hard let us, for the time being,
restrict ourselves to a rather special case of product basis set.
It is seen that by a suitable local unitary operator, $U_1\otimes
U_2$, it is always possible to transform two of the basis vectors
of $\{|\Psi_i\rangle\}_{i=1}^{4}$, say $|\Psi_1\rangle$ and
$|\Psi_2\rangle$, to the separable states $|00\rangle$ and
$|10\rangle$. If we choose
\begin{eqnarray}
&U_1=\left(\begin{array}{cc}\bar{a}&\bar{b}\\-b&a\end{array}\right),
\hskip 1mm
U_2=\left(\begin{array}{cc}\bar{c}&\bar{d}\\-d&c\end{array}\right),
\end{eqnarray}
then $U_1\otimes U_2$ acts on the vectors of
Eq.~(\ref{separablebasis}) as
\begin{eqnarray}
&&U_1 \otimes U_2 |\Psi_1\rangle=|00\rangle,\nonumber\\
&&U_1 \otimes U_2 |\Psi_2\rangle=|10\rangle,\nonumber\\
&&U_1 \otimes U_2 |\Psi_3\rangle=\left[(\bar{a}e+\bar{b}f)|0\rangle+(-be+af)|1\rangle\right]|1\rangle,\nonumber\\
&&U_1 \otimes
U_2|\Psi_4\rangle=\left[-(-\bar{b}\bar{e}+\bar{a}\bar{f})|0\rangle+(b\bar{f}+a\bar{e})|1\rangle\right]|1\rangle.\nonumber\\
\label{septr}
\end{eqnarray}
Thus this separable basis is locally equivalent to the following
more simpler set
\begin{eqnarray}
&\{|00\rangle, |10\rangle, (A|0\rangle+B|1\rangle)|1\rangle,
(-\bar{B}|0\rangle+\bar{A}|1\rangle)|1\rangle\},
\label{newsepbasis}
\end{eqnarray}
for the specific values of $A$ and $B$ as in Eq.~(\ref{septr}).
Here a point is in order. However for any unitary operators
$U_1,U_2\in$ SU(2) we have $E(U_1\otimes U_2
U|\Psi\rangle)=E(U|\Psi\rangle)$, generally it cannot be concluded
that $E(U U_1\otimes U_2 |\Psi\rangle)= E(U|\Psi\rangle)$. As an
example, {\textsc{ cnot}}$|00\rangle=|00\rangle$, but
{\textsc{cnot}}$(H\otimes
1)|00\rangle=\frac{|00\rangle+|11\rangle}{\sqrt{2}}$ ($H$ is the
Hadamard matrix). This point hints us that our solution(s) may not
be the most general one(s), that is, it is still possible to find
other SPE in addition to what we find below. We return to this
point a bit later. Anyway, action of a ${\cal U}$ on the basis
vectors in Eq.~(\ref{newsepbasis}), respectively, gives the
following results {\small{
\begin{eqnarray*}
&\hskip
-2mm\left(\begin{array}{c}e^{-ic_3}c^-\\0\\0\\-ie^{-ic_3}s^-\end{array}\right)\hskip
-2mm\left(\begin{array}{c}0\\-ie^{ic_3}s^+\\e^{ic_3}c^+\\0\end{array}\right)\hskip
-2mm\left(\begin{array}{c}-ie^{-ic_3}s^-B\\e^{ic_3}c^+A\\-ie^{ic_3}s^+A\\e^{-ic_3}c^-B\end{array}\right)\hskip
-2mm\left(\begin{array}{c}-ie^{-ic_3}s^-\bar{A}\\-e^{ic_3}c^+\bar{B}\\ie^{ic_3}s^+\bar{B}\\e^{-ic_3}c^-\bar{A}\end{array}\right).
\end{eqnarray*}}}
The maximal entanglement condition for these states requires that
\begin{eqnarray}
&|s^-c^-|=|s^+c^+|=\frac{1}{2},\nonumber\\
&|e^{2ic_3}c^+s^+A^2-e^{-2ic_3}c^-s^-B^2|=\nonumber\\
&|e^{2ic_3}c^-s^-A^2-e^{-2ic_3}c^+s^+B^2|=\frac{1}{2}.
\end{eqnarray}
As is seen the conditions on $c_1$ and $c_2$ are independent of
the the last conditions. Combination of this set of equations
gives the following results
\begin{eqnarray}
&\sin 2(c_1-c_2)=\pm 1,~~\sin 2(c_1+c_2)=\pm 1,\label{cond1}\\
&|e^{2ic_3}(\pm 1)A^2-e^{-2ic_3}(\pm 1)B^2|=1.\label{cond2}
\end{eqnarray}
By using the triangular inequality, $|z_1+z_2|\leq |z_1|+|z_2|$,
for two complex variables $z_i=|z_i|e^{i\phi_i}$, and noting that
the equality holds when $\phi_1=\phi_2 ({\text{mod}}~2\pi)$,
Eq.~(\ref{cond2}) can be simplified more. If we note that
$|e^{2ic_3}A^2\pm e^{-2ic_3}B^2|\leq|A|^2+|B|^2=1$, and
parameterize $A=\cos\theta$ and $B=e^{2i\phi}\sin \theta$
($2\phi$, as the phase, is chosen for later convenience), then
$|e^{2ic_3}\cos^2 \theta\pm e^{-2i(c_3-2\phi)}\sin\theta|=1$
requires that
\begin{eqnarray*}
&e^{2ic_3}=\pm e^{-2i(c_3-2\phi)},
\end{eqnarray*}
which has the following simple solutions
\begin{eqnarray*}
&c_3=\phi+\frac{k\pi}{4},~c_3=\phi+\frac{(2k+1)\pi}{4};~k\in
\mathbb{Z}.
\end{eqnarray*}
 Now if we use the freedoms in choosing $c_i$'s for a
class (Appendix \ref{appa}), there only remain the following
solutions
\begin{eqnarray}
&c_3=\phi, \frac{\pi}{4}+\phi. \label{phis}
\end{eqnarray}
We should now go back to the condition (\ref{cond1}). Apparently
four different cases can be obtained from this equation. But again
by a simple algebra and using the freedoms in $c_i$'s of a class
it appears that the two classes
\begin{eqnarray}
&{\cal U}_1\equiv [0,\frac{\pi}{4},\phi],\label{U1}\\
&{\cal U}_2\equiv [\frac{\pi}{4},0,\frac{\pi}{4}+\phi],\label{U2}
\end{eqnarray}
are our independent solutions. Furthermore, by a simple
redefinition of $\phi$ in Eq.~(\ref{U2}) and another use of the
freedom for $c_i$'s, it can be inferred that we actually have
found a one-parameter family of SPE
\begin{eqnarray}
&[\frac{\pi}{4},\phi,0],~~0\leq \phi\leq \frac{\pi}{4}.\label{SPE}
\end{eqnarray}
\begin{figure}[tp]
\psfrag{n}[Bc][][0.75][0]{$\frac{\pi}{4}$}
\psfrag{O}[Bc][][0.75][0]{$O$} \psfrag{c}[Bc][][0.75][0]{$c_1$}
\psfrag{a}[Bc][][0.75][0]{$A_1$} \psfrag{b}[Bc][][0.75][0]{$B$}
\psfrag{d}[Bc][][0.75][0]{$A_2$} \psfrag{e}[Bc][][0.75][0]{$c_2$}
\psfrag{f}[Bc][][0.75][0]{$c_3$}
\includegraphics[width=3.5cm,height=3.5cm]{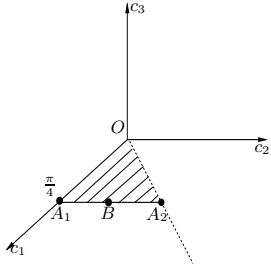}
\caption{Geometrical representation of SPE gates on the
$(c_1,c_2,c_3)$ space. The line $A_1A_2$ is the location of all
possible SPE. The points $A_1, B$, and $A_2$ denote the
[\textsc{cnot}], the $[B]$, and [\textsc{dcnot}] gates,
respectively. The filled area is projection of the Weyl chamber
which is the region containing nonlocal gates. } \label{fig1}
\end{figure}
Figure~\ref{fig1} shows the geometrical representation of the SPE
family in the $(c_1,c_2,c_3)$ space\footnote{The SPE line $A_1A_2$
in Fig.~\ref{fig1} corresponds to the $A_2L$ line in Fig.~1 of
\cite{Zhang4}.}. It should be noted that in the two-qubit case we
are fortunate to have this simple geometrical interpretation,
which is relatively absent for higher cases.

 Using the form, for
example, in Eq.~(\ref{U1}) we can simply find the form of the
transformed states (up to local unitary operators), which are as
\begin{eqnarray}
&e^{-i\phi}\frac{|00\rangle+i|11\rangle}{\sqrt{2}},\hskip 6mm e^{i\phi}\frac{-i|01\rangle+|10\rangle}{\sqrt{2}},\nonumber\\
& \frac{e^{i\phi}}{\sqrt{2}}[\sin
\theta(i|00\rangle+|11\rangle)+\cos
\theta(|01\rangle-i|10\rangle)],\nonumber\\
&\frac{e^{-i\phi}}{\sqrt{2}}[\cos
\theta(i|00\rangle+|11\rangle)+\sin
\theta(i|10\rangle-|01\rangle)], \label{actionsecond}
\end{eqnarray}
and are clearly maximally entangled. \\
It is interesting to see that the [{\textsc{cnot}}] gate (for
$\phi=0$) and the [{\textsc{dcnot}}] gate (for
$\phi=\frac{\pi}{4}$) are both members of this one-parameter
family of SPE, however, $[\sqrt{\rm{\textsc{swap}}}]$ is not. Also
the local invariants of the SPE in Eq.~(\ref{SPE}) are
\begin{eqnarray}
&G_1=0,~~~G_2=\cos 4\phi.
\end{eqnarray}

Up to now, we just have shown that at least there exists a
one-parameter family of unitary operators that are SPE,
Eq.~(\ref{SPE}). Now, we consider the problem in a general manner.
In fact, we show that the family (\ref{SPE}) is the only possible
family of operators that are SPE. To this end, we use the diagonal
form of ${\cal U}$ in the magic basis, Eq.~(\ref{spectral}). In
this basis we can write
\begin{eqnarray*}
&|\Psi_{\text{sep}}^{(i)}\rangle=\sum_k\mu_k^{(i)}|\Phi_k\rangle,\hskip
4mm|\Psi_{\text{me}}^{(i)}\rangle=\sum_k w_k^{(i)}|\Phi_k\rangle,
\end{eqnarray*}
where $\{|\Psi_{\text{sep}}^{(i)}\rangle\}_{i=1}^{4}$ is a general
product (separable) basis, and similarly
$\{|\Psi_{\text{me}}^{(i)}\rangle\}_{i=1}^{4}$ is a maximally
entangled basis. So our question, now, is to find unitary
operators, ${\cal U}$, for which we can find some suitable
separable basis ($\mu_k^{(i)}$'s) and a maximally entangled basis
($w_k^{(i)}$) such that
\begin{eqnarray}
{\cal
U}|\Psi_{\text{sep}}^{(i)}\rangle=|\Psi_{\text{me}}^{(i)}\rangle.
\label{upsi}
\end{eqnarray}
By using Eq.~(\ref{spectral}), the following relation between
$\mu_k^{(i)}$'s and $w_k^{(i)}$'s is obtained
\begin{eqnarray}
&w_k^{(i)}=e^{-i\lambda_k}\mu_k^{(i)}~~~~(i,k=1,\ldots,4).
\label{wmu}
\end{eqnarray}
In addition, orthonormality conditions,
$\langle\Psi_{\text{sep}}^{(i)}|\Psi_{\text{sep}}^{(j)}\rangle=\langle\Psi_{\text{me}}^{(i)}|\Psi_{\text{me}}^{(j)}\rangle=\delta_{ij}$
read as
\begin{eqnarray}
&\sum_k\overline{\mu_k^{(i)}}\mu_k^{(j)}=\sum_k\overline{w_k^{(i)}}w_k^{(j)}=\delta_{ij}.
\label{orth}
\end{eqnarray}
Now, taking into account the last points of Sec.~\ref{sec2} ,
separability of $\{|\Psi_{\text{sep}}^{(i)}\rangle\}_{i=1}^{4}$
and maximally entanglement of
$\{|\Psi_{\text{me}}^{(i)}\rangle\}_{i=1}^{4}$, respectively,
requires that
\begin{eqnarray}
&\sum_k\mu_k^{(i)}=0,\label{4}\\
&w_k^{(i)~2}=e^{-i\delta^{(i)}}|w_k^{(i)}|^2\label{5},
\end{eqnarray}
from which (by using Eq.~(\ref{wmu})) one obtains
\begin{eqnarray}
&\sum_k e^{2i\lambda_k}w_k^{(i)~2}=0,\label{6}\\
&w_k^{(i)}=\pm e^{\frac{-i}{2}\delta^{(i)}}|w_k^{(i)}|.\label{7}
\end{eqnarray}
If we replace Eq.~(\ref{7}) in Eq.~(\ref{6}) and multiply it by
$e^{2ic_3}$, after separating real and imaginary parts we get
\begin{eqnarray}
&\hskip -15mm|w_3^{(i)}|^2+|w_1^{(i)}|^2\cos\alpha_2+|w_2^{(i)}|^2\cos\alpha_3+|w_4^{(i)}|^2\cos\alpha_1=0,\label{8}\\
&|w_1^{(i)}|^2\sin\alpha_2+|w_2^{(i)}|^2\sin\alpha_3+|w_4^{(i)}|^2\sin\alpha_1=0,\label{9}
\end{eqnarray}
for $i=1,\ldots,4$, where\\
\begin{eqnarray*}
&\alpha_1=4(c_1+c_2),~\alpha_2=4(c_1+c_3),~\alpha_3=4(c_2+c_3).
\end{eqnarray*}
Orthonormality condition, Eq.~(\ref{orth}), is equal to unitarity
of the matrix $U_{ik}:= w_k^{(i)}$, which instead gives the below
results
\begin{eqnarray}
&\sum_k|w_k^{(i)}|^2=\sum_i|w_k^{(i)}|^2=1,\label{10}\\
&\sum_k\overline{w_k^{(i)}}w_k^{(j)}=\sum_k\overline{w_i^{(k)}}w_j^{(k)}=\delta_{ij}.\label{11}
\end{eqnarray}
By using Eq.~(\ref{7}), Eq.~(\ref{11}) can be simplified as
\begin{eqnarray}
\hskip
-3mm&e^{\frac{i}{2}(\delta^{(i)}-\delta^{(j)})}\sum_k|w_k^{(i)}||w_k^{(j)}|=\sum_k|w_i^{(k)}||w_j^{(k)}|=\delta_{ij}.\label{11-1}
\end{eqnarray}
If we now add the relations in Eq.~(\ref{8}) for different $i$'s,
and consider Eq.~(\ref{10}), it is obtained that
\begin{eqnarray}
&\cos\alpha_1+\cos\alpha_2+\cos\alpha_3=-1.\label{12}
\end{eqnarray}
In a similar manner, Eq.~(\ref{9}) results into
\begin{eqnarray}
&\sin\alpha_1+\sin\alpha_2+\sin\alpha_3=0.\label{13}
\end{eqnarray}
By combination of these two equations, the following final
condition for a class $[c_1,c_2,c_3]$ being a SPE is obtained
\begin{eqnarray}
&\hskip -6mm\cos4c_1 \cos4c_2+\cos4c_2 \cos4c_3+\cos4c_3
\cos4c_1=-1.\label{epcond}
\end{eqnarray}
This, in turn, implies that $e_p({\cal U})=\frac{2}{9}$.
Therefore, we have proved that if a unitary operator, ${\cal U}$,
is a SPE it must have the maximal entangling power. Now, let us
investigate all possible solutions of Eq.~(\ref{epcond}). It is
trivial to check that if one of the $c_i$'s, say $c_1$, is equal
to $\frac{\pi}{4} (0)$, then the only choices for the other
$c_i$'s are $0 (\frac{\pi}{4})$ and $0\leq \phi \leq
\frac{\pi}{4}$. This is the very family of SPE as in
Eq.~(\ref{SPE}). More generally, the only acceptable solutions of
Eq.~(\ref{epcond}) are also of the form in (\ref{SPE}) (and its
permutations).\\
Thus it has been shown that only the unitary operators
$[\frac{\pi}{4},\phi,0]$ which are the unique family with
$e_p=\frac{2}{9}$ can be SPE. This fact stresses that for the SPE
entangling power is a sufficient tool which can characterize them
uniquely.

 Now let us consider the subspace of non-perfect
entanglers, especially the [{\textsc{swap}}] class in the sense
that how they act on separable bases. We want to point out a
simple property of this class. As has been proved in
\cite{Brylinski} any nonlocal two-qubit (qudit) unitary operator
other than those in the [{\textsc{swap}}] class can be used as a
universal operator (of course together with elementary
single-qubit unitary operators). Additionally, it has been known
that any unitary operator $V$ which leaves product states product
(that is, for any $|x\rangle$ and $|y\rangle$ there exist some
$|u\rangle$ and $|v\rangle$ such that
$V|x\rangle|y\rangle=|u\rangle|v\rangle$) is either a local
unitary operator $A\otimes B$ or $(A\otimes B) S$, for some $A$
and $B$. Now, it can be deduced that the only unitary operators
that can transform {\em any} separable basis to some separable
basis are those in the [{\textsc{swap}}] class. This can be seen
as follows. Consider two arbitrary vectors $|a\rangle$ and
$|b\rangle$. Suppose that $V$ is such an operator that transforms
any product basis to another one. To see that if
$V|a\rangle|b\rangle$ is a separable state, we can construct
another basis set in which $|a\rangle|b\rangle$ is one of the
basis vectors. For example,
\begin{eqnarray*}
&\{|a\rangle|b\rangle,|a^{\perp}\rangle|b\rangle,|c\rangle|b^{\perp}\rangle,|c^{\perp}\rangle|b^{\perp}\rangle\},
\end{eqnarray*}
is such a new basis. Thus $V$ must transform $|a\rangle|b\rangle$
to some $|a'\rangle|b'\rangle$. As mentioned above, this indicates
that $V$ is either local or locally equivalent to the
{\textsc{swap}} operator. Here we should remark that the above
statement is true only if $V$ leaves any separable basis a
separable one. For example, the {\textsc{ cnot}} operator
transforms the usual computational basis
$\{|00\rangle,|01\rangle,|10\rangle,|11\rangle\}$ to another
separable basis, however, it is nonlocal.

Finally we show that by using the spectral decomposition of the
nonlocal part, Eq.~(\ref{spectral}), a simple
(yet remarkable) property of  ${\cal U}$, which is related to the magic basis, can be deduced.\\
\indent {\em Lemma.}~ There is no separable basis
$\{|\alpha_i\rangle\otimes|\beta_i\rangle\}_{i=1}^{4}$ that can be
transformed to the magic basis by an operator  ${\cal U}$, as in Eq.~(\ref{explicitU}).\\
\indent {\em Proof}: (By contradiction)~Suppose that for a given
operator  ${\cal U}$ there exists such a basis, ${\cal
U}|\alpha_i\rangle\otimes |\beta_i\rangle=|\Phi_i\rangle$. This is
equivalent to ${\cal
U}^{\dagger}|\Phi_i\rangle=|\alpha_i\rangle\otimes|\beta_i\rangle$.
By using the form in Eq.~(\ref{spectral}), this gives rise to
$e^{i\lambda_i}|\Phi_i\rangle=|\alpha_i\rangle\otimes|\beta_i\rangle$
which is an evident contradiction!  ~~~~~~$\Box$\\
It should be emphasized that, for example, for the
[{\textsc{cnot}}] class we have
\begin{eqnarray*}
&\frac{|0\rangle \pm |1\rangle}{\sqrt{2}}|0\rangle\rightarrow
\frac{|00\rangle \pm |11\rangle}{\sqrt{2}},\\
&\frac{|0\rangle \pm |1\rangle}{\sqrt{2}}|1\rangle\rightarrow
\frac{|01\rangle \pm |10\rangle}{\sqrt{2}}.
\end{eqnarray*}
The right hand side is the usual Bell basis. However, this does
not violate the above lemma. In fact, the point is the difference
between the usual Bell basis and the magic basis. The two sets of
basis vectors are not locally related. This can be seen by
examining, for example, the following operator
\begin{eqnarray*}
&U=|\Phi_1\rangle\langle\Phi^+|+|\Phi_2\rangle\langle\Phi^-|+|\Phi_3\rangle\langle\Psi^-|+|\Phi_4\rangle\langle\Psi^+|.
\end{eqnarray*}
This operator is diagonal in the magic basis, and thus is
nonlocal. Actually, for this specific operator $G_1[U]=0$ and
$G_2[U]=-1$ which indicate $U\in$ [{\textsc{dcnot}}].

\section{Applications to efficient two-qubit gate simulation}\label{sec4}
In this section, we are going to present a remarkable importance
of the SPE, which can put forward special theoretical and
experimental interests in them.

 From theoretical and experimental viewpoints,
efficient quantum gate synthesis is of high interest. In fact,
there is a vast literature on this topic with many remarkable
findings \cite{Brylinski,Barenco, Barencoetal, Lloyd, Bremner,
Zhang2, Shende1, Shende2,Bullock, Vidaldawson, Vatan, Zhang3,
Zhang4, Vart}. It has been known that a general (multi-qubit)
quantum gate can be simulated using a quantum circuit built of
elementary gates which operate on single and two qubits
\cite{Barenco}. For instance, Barenco {\em et al.} \cite{Barenco}
have shown that a combination of the \textsc{cnot} and
single-qubit gates is universal, in the sense that any unitary
operation can be simulated by them. This is the commonly adopted
universal set in paradigm of quantum computation. However, there
has been a lot of interest in universal sets which are highly (or
the most) efficient. Actually, there is a practical reason behind
this demand. In present day experiments, two-qubit gates as the
\textsc{cnot} gate can only be implemented imperfectly due to
today technological limitations (the so-called decoherence).
Therefore, in order to reduce the probability that an error occurs
in performing a certain unitary operation on several qubits,
instrumentally it is highly demanded that the number of the
building block gates (i.e., the number of times that the qubits
interact) be as small as possible. Furthermore, it is clear that
these attempts can be of advantage in analyzing the algorithmic
complexity of a given quantum computation and other QIP related
tasks \cite{Nielsen}.

Very recently, it has been shown that in order to implement an
arbitrary two-qubit unitary operator not more than three
applications of the \textsc{cnot} or the \textsc{dcnot} gates are
necessary \cite{Vidaldawson, Zhang3}. In this sense, the
\textsc{cnot} and the \textsc{dcnot} gates are highly efficient
elementary gates. However, remarkably, it has also recently been
shown that the
$B=e^{-i(\frac{\pi}{4}\sigma_1\otimes\sigma_1+\frac{\pi}{8}\sigma_2\otimes\sigma_2)}$
gate is the most efficient one, in that any arbitrary two-qubit
unitary operator can be constructed only by two applications of
such gate \cite{Zhang4}. The fact we would like to stress on is
that all of these three gates; the \textsc{cnot}, the
\textsc{dcnot}, and the $B$ gates, are SPE. This can raise the
question that if other SPE gates also can give rise to a minimum
universal gate construction. To this end, we follow the method
proposed in \cite{Zhang3,Zhang4}. We need to remind that the $B$
gate as the most efficient known universal gate can simulate a
generic nonlocal two-qubit operator with the following circuit
\begin{eqnarray}
\label{Bgate}
%\begin{widetext}
\setlength{\unitlength}{0.15cm} \scriptsize
\begin{picture}(48,8)
%\put(0,0){\framebox(70,7)}
\put(1,1.7){\line(1,0){2}} \put(1,5.5){\line(1,0){2}}
\put(3,0){\thicklines\framebox(4,7)[c]{$B$}}
\put(7,1.5){\line(1,0){3}} \put(7,5.5){\line(1,0){8}}
\put(10,0){\framebox(17,3)[c]{$e^{-ia\sigma_3}e^{-ib\sigma_2}e^{-ia\sigma_3}$}}
\put(15,4){\framebox(7,3)[c]{$e^{-ic_1\sigma_2}$}}
\put(27,1.5){\line(1,0){3}} \put(22,5.5){\line(1,0){8}}
\put(30,0){\thicklines\framebox(4,7)[c]{$B$}}
\put(34,1.5){\line(1,0){2}} \put(34,5.5){\line(1,0){2}}
\put(37,3){$\equiv$}\put(39,1.7){\line(1,0){2}}\put(39,5.5){\line(1,0){2}}\put(41,0){\thicklines\framebox(4,7)[c]{${\cal
U}$}}\put(45,1.7){\line(1,0){2}}\put(45,5.5){\line(1,0){2}}
\end{picture}
%\end{widetext}
\end{eqnarray}
where the parameters $a$ and $b$ satisfy
\begin{eqnarray}
\label{constr} &\sin 2a =\sqrt{\frac{\cos 2c_2 \cos 2c_3}{1-2
\sin^2 c_2 \cos^2 c_3}},\nonumber\\
&\cos 2b=1-4 \sin^2 c_2 \cos^2 c_3.
\end{eqnarray}
Equivalence of these two circuits can be proved simply by showing
that the nonlocal invariants ($G_1$ and $G_2$) of the both
circuits are the same. For our study, we choose an arbitrary SPE,
which we call
$C[\phi]=e^{-i(\frac{\pi}{4}\sigma_1\otimes\sigma_1+\phi\sigma_2\otimes\sigma_2)}$.
 We are looking for some SPE gates other than the
$B=C[\frac{\pi}{8}]$ gate that give rise to a minimum construction
of two-qubit unitary operators universally. Following
\cite{Zhang4}, we consider the circuit below which is similar to
the one in Eq.~(\ref{Bgate})
\begin{center}
\begin{eqnarray}
\label{Cgate}
%\begin{widetext}
%\hskip 10mm
 \setlength{\unitlength}{0.15cm} \scriptsize
\begin{picture}(36,3)
%\put(0,0){\framebox(70,7)}
\put(1,1.7){\line(1,0){2}} \put(1,5.5){\line(1,0){2}}
\put(3,0){\thicklines\framebox(4,7)[c]{$C[\phi]$}}
\put(7,1.5){\line(1,0){3}} \put(7,5.5){\line(1,0){8}}
\put(10,0){\framebox(17,3)[c]{$e^{-ia\sigma_3}e^{-ib\sigma_2}e^{-ia\sigma_3}$}}
\put(15,4){\framebox(7,3)[c]{$e^{-ic_1\sigma_2}$}}
\put(27,1.5){\line(1,0){3}} \put(22,5.5){\line(1,0){8}}
\put(30,0){\thicklines\framebox(4,7)[c]{$C[\phi]$}}
\put(34,1.5){\line(1,0){2}} \put(34,5.5){\line(1,0){2}}
\end{picture}
%\end{widetext}
\end{eqnarray}
\end{center}
and try to find the parameters $a$ and $b$ such that this circuit
can simulate a generic nonlocal gate ${\cal U}(c_1,c_2,c_3)$, as
in Eq.~(\ref{explicitU}). After computing the local invariants of
this circuit and comparing them with the invariants of
$[c_1,c_2,c_3]$, it is seen that the following equations must be
satisfied
\begin{widetext}
\begin{eqnarray}
\label{constraints} &\cos 2a \sin 2b\sin 4\phi=\sin 2c_2 \sin
2c_3\\
&(1+\cos 4\phi)\cos 4a+(1+\cos 4\phi)\cos 4a\cos 2b+(3\cos
4\phi-1)\cos 2b+4\cos 2c_2\cos 2c_3-(1+\cos 4\phi)=0
\end{eqnarray}
\end{widetext}
In fact, there is another (relatively long) equation which also
must be satisfied as the consistency check for the possible
solutions, which we have not written here. By the way, it is easy
to see that the following two solutions can be found
%\begin{widetext}
\begin{eqnarray}
\label{sols1}
 &\cos 2b=1-\frac{(1-\cos 2c_3)(1+\cos 2c_2)}{1-\cos
4\phi},\nonumber\\
&\cos^2 2a=\frac{(1+\cos 2c_3)(1-\cos 2c_2)\tan^2
2\phi}{1+\cos2c_3-\cos 2c_2-\cos 2c_2\cos 2c_3},
\end{eqnarray}
%\end{widetext}
and
\begin{eqnarray}
\label{sols2}
 &\cos 2b=1-\frac{(1+\cos 2c_3)(1-\cos 2c_2)}{1-\cos
4\phi},\nonumber\\
&\cos^2 2a=\frac{(1-\cos 2c_3)(1+\cos 2c_2)\tan^2
2\phi}{1-\cos2c_3+\cos 2c_2+\cos 2c_2\cos 2c_3}.
\end{eqnarray}
Both of these solutions satisfy the mentioned consistency check
relation, for any $\phi$. However, from these solutions it
immediately can be deduced that we must have $\phi\neq 0,
\frac{\pi}{4}$. This means that the \textsc{cnot} and the
\textsc{dcnot} gates cannot be among these gates. This fact is in
accordance to the results of \cite{Vidaldawson, Zhang3} which
indicate that except some special cases at least three
applications of the \textsc{cnot} and the \textsc{dcnot} gates are
necessary and sufficient for generic gate simulations. In
addition, after checking that the solutions must satisfy the the
relations $-1\leq \cos 2b\leq 1$ and $0\leq \cos^2 2a\leq 1$, it
is found that if we want to have at least one solution we must
choose the value of $\phi$ by considering the values of $c_2$ and
$c_3$ of ${\cal U}(c_1,c_2,c_3)$ operator which is to be
simulated. That is, not any $\phi$ (and hence $C[\phi]$) is proper
for simulation of an arbitrary nonlocal gate. It can directly be
checked that the only value of $\phi$ which is applicable in
simulation of {\em all} two-qubit unitary operators is
$\phi=\frac{\pi}{8}$. Thus, in this way, we argue that the most
efficient gate that can be used in universal two-qubit gate
construction is the $B=C[\frac{\pi}{4}]$ gate. This fact gives a
unique and special role to the $B$ gate among other SPE
\cite{Zhang4}.

 To complete our discussion, in the following we show how one can
simulate some well-known gates, such as the \textsc{cnot} and the
\textsc{dcnot}, using our method. For the case of the
\textsc{cnot} gate, we have the following equivalent circuit
\begin{eqnarray*}
 \setlength{\unitlength}{0.15cm} \scriptsize
\begin{picture}(36,7)
%\put(0,0){\framebox(70,7)}
\put(1,1.7){\line(1,0){2}} \put(1,5.5){\line(1,0){2}}
\put(3,0){\thicklines\framebox(4,7)[c]{$C[\phi]$}}
\put(7,1.5){\line(1,0){11}} \put(7,5.5){\line(1,0){2}}
\put(9,4){\framebox(7,3)[c]{$e^{-i\frac{\pi}{4}\sigma_2}$}}
\put(16,5.5){\line(1,0){2}}
\put(18,0){\thicklines\framebox(4,7)[c]{$C[\phi]$}}
\put(22,1.5){\line(1,0){2}} \put(22,5.5){\line(1,0){2}}
\put(26,3){$\equiv$}\put(29,1.5){\line(1,0){2}}
\put(29,5.5){\line(1,0){2}}\put(31,0){\thicklines\framebox(5,7)[c]{\textsc{cnot}}}\put(36,1.5){\line(1,0){2}}
\put(36,5.5){\line(1,0){2}}
\end{picture}
\end{eqnarray*}
which is valid for $0<\phi<\frac{\pi}{4}$. Similarly, for the
\textsc{dcnot} gate, the following circuit gives an equivalent
construction
\begin{eqnarray*}
\setlength{\unitlength}{0.15cm} \scriptsize
\begin{picture}(48,8)
%\put(0,0){\framebox(70,7)}
\put(1,1.7){\line(1,0){2}} \put(1,5.5){\line(1,0){2}}
\put(3,0){\thicklines\framebox(4,7)[c]{$C[\phi]$}}
\put(7,1.5){\line(1,0){2}} \put(7,5.5){\line(1,0){8}}
\put(9,0){\framebox(19,3)[c]{$e^{-i\frac{\pi}{4}\sigma_3}e^{-ib\sigma_2}e^{-i\frac{\pi}{4}\sigma_3}$}}
\put(15,4){\framebox(7,3)[c]{$e^{-i\frac{\pi}{4}\sigma_2}$}}
\put(28,1.5){\line(1,0){2}} \put(22,5.5){\line(1,0){8}}
\put(30,0){\thicklines\framebox(4,7)[c]{$C[\phi]$}}
\put(34,1.5){\line(1,0){2}} \put(34,5.5){\line(1,0){2}}
\put(37,3){$\equiv$}\put(39,1.7){\line(1,0){2}}\put(39,5.5){\line(1,0){2}}\put(41,0){\thicklines\framebox(6,7)[c]{\textsc{dcnot}}}
\put(47,1.7){\line(1,0){2}}\put(47,5.5){\line(1,0){2}}
\end{picture}
\end{eqnarray*}
where
\begin{eqnarray*}
&\cos 2b=-\cot^2 2\phi,~~~\frac{\pi}{8}\leq\phi<\frac{\pi}{4}.
\end{eqnarray*}

As the final point, we want to emphasize that how the $C[\phi]$
gates can be simulated in the standard gate construction, i.e., by
applications of the \textsc{cnot} (or \textsc{dcnot}) gate. In
\cite{Vidaldawson, Zhang3}, it has been shown that any nonlocal
gate as $[c_1,c_2,0]$, where $\frac{\pi}{4}\geq c_1\geq c_2 \geq
0$ and $c_1>0$ (except for $[\frac{\pi}{4},0,0]$), can be
simulated with only two \textsc{cnot} or \textsc{dcnot} gates.
This means that all SPE, the $C[\phi]$ gates
($0<\phi<\frac{\pi}{4}$), can also be constructed very efficiently
using the \textsc{cnot} or the \textsc{dcnot} gates.

In summary, what we have shown is that for optimal construction of
any arbitrary two-qubit gate  we can always find (at least) one
circuit synthesis by some suitable SPE. In our analytic
construction, the choice of the convenient SPE is governed by the
values of $c_2$ and $c_3$ of the gate which is to be simulated.
From these proper SPE one can pick the one that is available or
can be implemented easily in experiment. In this way, typically
there are more options at hand that one can utilize in
experimental implementations. In addition, we have clarified the
unique property of the $B$ gate, that is it poses the most
efficient elementary gate suitable in all generic two-qubit gate
constructions.

\section{Conclusions}{\label{conc}}
It has been shown that some of perfect entanglers have the unique
property that they can maximally entangle a suitable full
separable basis. Some well-known two-qubit gates such as the
{\textsc{cnot}} and the {\textsc{dcnot}} are among such special
operators, however $\sqrt{\rm{\textsc{swap}}}$ is not. In our case
of study entangling power has appeared as a useful tool in
characterization of special perfect entanglers. The perfect
entanglers with maximum value of entangling power, $\frac{2}{9}$,
have turned out to be the only possible special perfect
entanglers.

 As a possible application of these special gates their
importance in universal two-qubit gate construction has been
emphasized. We have provided a more general (analytic) result for
optimality. In this regard, it has been proved that if we have the
SPE gates at our disposal we can simulate any generic two-qubit
gate by only two applications of the same gates. In fact, in this
manner we have a vast possibility for universal gate construction.
Although not any SPE gate may be applicable in simulation of a
specific two-qubit gate, it is always possible to find some proper
SPE for this mean. Specially, the unique role of the $B$ gate,
namely, its usage in all two-qubit gate simulations, has been
pointed out. In this sense, these results can shed new light on
theoretical and experimental investigations to characterize the
rich structure of the space of nonlocal unitary operators and
using them as the most efficient universal elementary gates.
\begin{acknowledgments}
The author would like to thank V. Karimipour for useful comments.
\end{acknowledgments}

\appendix
\section{Possible freedoms in nonlocal parts}\label{appa}
It can be simply checked that in a triplet $[c_1,c_2,c_3]$, which
represents nonlocal part of an operator, one can easily add any
integer multiple of $\frac{\pi}{2}$ to every $c_i$. This can be
verified by multiplication of the local operators $\pm i\sigma_i
\otimes \sigma_i=e^{\pm i\frac{\pi}{2}\sigma_i \otimes \sigma_i}$.
Also $[c_1,c_2,c_3]=[c_i,c_j,c_k]$, where $(ijk)$ is an arbitrary
permutation of $(123)$. This can be verified as follows. If in the
identity $e^{u_1\otimes v_1 V u_1^{\dag}\otimes
v_1^{\dag}}=u_1\otimes v_1 e^{V}u_1^{\dag}\otimes v_1^{\dag}$,
where all operators are unitary, we take $u_1\otimes
v_1=e^{-i\frac{\pi}{4}\sigma_1}\otimes
e^{-i\frac{\pi}{4}\sigma_1}$ it is obtained that
$[c_1,c_2,c_3]=[c_1,c_3,c_2]$. A similar procedure can be used to
prove the general statement. The other fact about $[c_1,c_2,c_3]$
is that one can always make at least two of $c_i$'s positive (or
negative). This can be verified by taking $u_1\otimes v_1 =\pm
i\sigma_k\otimes 1$ in the above mentioned identity. To see more
on the freedoms refer, for example, to \cite{Hammerer}.

\section{The local invariants}\label{appb}
Consider the matrix $Q$ which is
\begin{eqnarray}
&Q=\frac{1}{\sqrt{2}}\left(\begin{array}{cccc}1&0&0&i\\0&i&1&0\\0&i&-1&0\\1&0&0&-i\end{array}\right).
\end{eqnarray}
Then define $U_B:=Q^{\dag}UQ$, and accordingly $m(U):=U_B^{T}U_B$
($T$ means transposition). $G_1[U]$ and $G_2[U]$ defined as
\begin{eqnarray}
&&G_1[U]:=\frac{{\text{tr}}^2 [m(U)]}{16~{\text{det}}~U},\\
&&G_2[U]:=\frac{{\text{tr}}^2 [m(U)]-{\text{tr}}[m^2(U)]}{4~
{\text{det}}~U}, \label{invariant}
\end{eqnarray}
have been proved to be locally invariant \cite{Makhlin, Zhang1}.
Therefore they can characterize equivalence classes. In this
respect we can write $G_i[U]=G_i[c_1,c_2,c_3]$. For a nonlocal
operator, as in Eq.~(\ref{explicitU}), a simple algebra reveals
that
\begin{eqnarray}
&&\hskip -6mmG_1=\frac{1}{4}[e^{-2i c_3}\cos 2(c_1-c_2)+e^{2i
c_3}\cos
2(c_1+c_2)]^2,\hskip 4mm\label{invs11}\\
&&\hskip -6mmG_2=\cos (4 c_1)+\cos (4 c_2)+\cos (4 c_3).
\label{invs22}
\end{eqnarray}
Table~\ref{tab1} summarizes the results for some examples. As is
seen, the [{\textsc{cnot}}], [{\textsc{dcnot}}], and
[$\sqrt{\rm{\textsc{swap}}}$] all belong to different equivalence
classes.
\begingroup
\squeezetable
\begin{table}[tp]
\begin{ruledtabular}
\caption{Nonlocal properties of some equivalence classes.}
\begin{tabular}{ccccc}
 Operator & $[c_1,c_2,c_3]$ & $G_1$&$G_2$&$e_p${\footnotemark[1]}\\
\colrule
{\textsc{cnot}} & $[\frac{\pi}{4},0,0]$&0&1&$\frac{2}{9}$\\
{\textsc{dcnot}} & $[\frac{\pi}{4},\frac{\pi}{4},0]$ & 0&-1&$\frac{2}{9}$\\
$B$ & $[\frac{\pi}{4},\frac{\pi}{8},0]$ & 0&0&$\frac{2}{9}$\\
{\textsc{swap}} & $[\frac{\pi}{4},\frac{\pi}{4},\frac{\pi}{4}]$ &-1&-3&0\\
$\sqrt{\rm{\textsc{swap}}}$ & $[\frac{\pi}{8},\frac{\pi}{8},\frac{\pi}{8}]$ & $\frac{i}{4}$&0&$\frac{1}{6}$\\
controlled-$U$ & $[x,0,0]${\footnotemark[2]}&$G_1$ & $2G_1+1$&$\frac{1-\cos 4x}{9}$\\
$A\otimes B$&$[0,0,0]$&1&3&0\\
\end{tabular}\label{tab1}
\end{ruledtabular}
\footnotetext[1]{This is entangling power which is defined in
Eq.~(\ref{ep}).} \footnotetext[2]{The exact value of $x$ depends
on explicit form of $U$. For
example, if $U=$ {\tiny{$\left(\begin{array}{cc}\cos \theta & \sin\theta e^{-i\phi} \\
-\sin \theta e^{i\phi}& \cos \theta
\end{array}\right)$}}
 then $x=\frac{\theta}{2}$, and $G_1=\cos^{2}\theta
$.}
\end{table}
\endgroup
There is a powerful theorem which exactly determines
which unitary operators are perfect entangler \cite{Makhlin,Zhang1}.\\
\indent {\em Theorem}:~~ A two-qubit unitary operator is a perfect
entangler iff the convex hull of the eigenvalues of $m(U)$
contains zero.\\
%It is reminded that the convex hull ${\cal C}$ of $N$ points $p_1,
%p_2,\ldots,p_N \in \mathbb{R}^n$ is given by
%\begin{eqnarray}
%&{\cal C}=\{\sum_{i=1}^{N}\lambda_i p_i|\forall i;~0\leq
%\lambda_i\leq 1,~\sum_{i=1}^{N}\lambda_i=1\}.
%\end{eqnarray}
An immediate result of this theorem is that  among controlled-$U$
operators only the [{\textsc{cnot}}] class is perfect entangler
(Table~\ref{tab1}).

\section{Entangling power}\label{appc}
Entangling power, $e_p(U)$, is a widely accepted measure for
quantifying entangling capabilities of a unitary quantum operator
$U$ \cite{Zanardi1}. Consider a bipartite quantum system with
state space of ${\cal H}_{A}\otimes {\cal H}_{B}$ (in qubit case
${\cal H}_{A}={\cal H}_{B}=\mathbb{C}^2$). If $E$ is an
entanglement measure over ${\cal H}_{A}\otimes {\cal H}_{B}$, the
entangling power of $U$ (with respect to $E$) is defined as
\begin{eqnarray}
&e_p(U):=\overline{E(U|\psi\rangle\otimes
|\phi\rangle)}^{~|\psi\rangle,|\phi\rangle}, \label{ep}
\end{eqnarray}
where the bar denotes the average over all the product states
$|\psi\rangle\otimes|\phi\rangle$ distributed according to some
probability density $p(\psi,\phi)$ over the manifold of product
states. To ensure local invariance for $e_p(U)$ the measure of
integration, $p$, must be taken locally invariant. A simple and
proper distribution which satisfies this requirement is the
uniform  distribution. In \cite{Zanardi1} $E$ has been taken to be
 linear entropy, which is defined as
\begin{eqnarray}
&E(|\Psi\rangle_{AB}):=1-{\text{tr}}(\rho_{A(B)}^2),
\end{eqnarray}
where
$\rho_{A(B)}={\text{tr}}_{B(A)}(|\Psi\rangle_{AB}\langle\Psi|)$ is
the reduced density matrix of the system $A (B)$. Now it is in
order to mention the relation of $E$ and the other (rather more
well-known) measure of entanglement, concurrence \cite{Wootters}.
Concurrence of any two-qubit density matrix, $\rho_{AB}$, is
defined as
\begin{eqnarray}
&C(\rho_{AB}):={\text{max}}
\{0,\lambda_1-\lambda_2-\lambda_3-\lambda_4\}, \label{concurrence}
\end{eqnarray}
in which $\lambda_1\geq \lambda_2 \geq \lambda_3 \geq \lambda_4$
are square roots of eigenvalues of the matrix
$\rho_{AB}\sigma_2\otimes \sigma_2 \rho_{AB}^{*}\sigma_{2}\otimes
\sigma_{2}$ ($*$ means complex conjugation in the standard
computational basis). For a pure state of two qubits,
$|\Psi\rangle=a|00\rangle+b|01\rangle+c|10\rangle+d|11\rangle$,
this reduces to $C(\Psi)=2|ad-bc|$. In this case, one can see that
eigenvalues of the reduced density matrices are
$\lambda_{\pm}=\frac{1\pm\sqrt{1-C^2(\Psi)}}{2}$, from which
\begin{eqnarray}
&E(\Psi)=\frac{1}{2}C^2(\Psi). \label{EC}
\end{eqnarray}
Since $C$ ranges from $0$ (for separable states) to $1$ (for
maximally entangled states), we have $0\leq E\leq \frac{1}{2}$.

Some important properties of $e_p(U)$ are listed below.\\
\indent (i)~For every $U\in$ SU(4) we have $0\leq e_p(U)\leq
\frac{2}{9}$.\\
\indent (ii)~For every $A,B\in$ SU(2), $e_p(A\otimes B)=0$. This
is clear from the fact that $E(|\psi\rangle\otimes
|\phi\rangle)=0$.\\
\indent (iii)~$e_p(U)$ is locally invariant, that is, for every
$A,B\in$ SU(2), $e_p(A\otimes B U)=e_p(U A\otimes B)=e_p(U)$. This
property is also simple to verify by noting that
\begin{eqnarray*}
&\overline{E(U A|\psi\rangle\otimes
B|\phi\rangle)}^{~|\psi\rangle,|\phi\rangle}=\overline{E(U
|\psi'\rangle\otimes
|\phi'\rangle)}^{~|\psi'\rangle,|\phi'\rangle}.
\end{eqnarray*}
 \indent (iv)~For
every $U\in$ SU(4), $e_p(U^{\dag})=e_p(U)$. The
equality holds because of $E(|\Psi^*\rangle)=E(|\Psi\rangle)$.\\
\indent (v)~If $S$ is the usual {\textsc{swap}} operator
($S|\psi\rangle\otimes
|\phi\rangle=|\phi\rangle\otimes|\psi\rangle$, for every
$|\psi\rangle$ and $|\phi\rangle$), then $e_p(SU)=e_p(US)=e_p(U)$.
These are immediate consequences of
$E(S|\Psi\rangle)=E(|\Psi\rangle)$, for all $|\Psi\rangle$.\\
\indent (vi)~Another trivial property is that $e_p(S)=0$.\\
To see more about the entangling power, its properties, and
generalizations refer to \cite{Zanardi1, Wangzanardi,
Wangsanders}.\\
By using the above facts, one can find an analytical form for
entangling power of a unitary operator as in Eq.~(\ref{EPclosed}).
The last equality follows from Eq.~(\ref{EC}) and using the
parametrization
$|\psi\rangle=$ {\small{$\left(\begin{array}{c}\cos\frac{\theta}{2}e^{i\frac{\phi}{2}}\\
\sin\frac{\theta}{2}e^{-i\frac{\phi}{2}}\end{array}\right)$}}
($0\leq \theta\leq \pi,0\leq\phi<2\pi$) for a single-qubit state
vector. Then a simple integration as
$e_p(U)=\frac{1}{(4\pi)^2}\int\frac{C^2}{2}\sin\theta d\theta
d\phi \sin\theta' d\theta' d\phi'$ gives the result. This form for
$e_p$ is clearly locally invariant, and shows that all unitary
operators in the same equivalence class possess an equal
entangling power, however the converse is not true generally. It
is  simple to check that this expression respects all the freedoms
mentioned earlier for $c_i$'s of a class. The last column in
Table~\ref{tab1} gives $e_p$ for some well-known equivalence
classes. It is important to note that the single parameter $e_p$
cannot replace the two local invariants $G_1$ and $G_2$ to
uniquely characterize equivalence classes.
%%%%%%%%%%%%%%%%%%%%%%%%%%%%%%%%%%%%%%%%%%%%%%%%%%%%%%%%%%%%%%

%%%%%%%%%%%%%%%%%%%%%%%%%%%%%%%%%%%%%%%%%%%%%%%%%%%%
\end{document}